\documentclass{eptcs}
\usepackage{graphicx}
\usepackage{paralist}
\usepackage{breakurl}
\usepackage{color}
\definecolor{lstbg}{rgb}{0.98,0.98,0.98}

\usepackage{listings}
\usepackage[normalem]{ulem}

\makeatletter
\lst@AddToHook{TextStyle}{\let\lst@basicstyle\normalsize\sffamily}
\makeatother

\newcommand{\smme}[1]{\uline{\texttt{#1}}}
\newcommand{\tmme}[1]{\texttt{#1}}

\newcommand{\myemail}[0]{\email{horn@uni-koblenz.de}}

\def\titlerunning{Program Understanding: A Reengineering Case for the
  Transformation Tool Contest}
\def\authorrunning{Tassilo Horn}

\title{Program Understanding: A Reengineering Case for the Transformation Tool
  Contest}
\author{Dipl.-Inform. Tassilo Horn\\
  \myemail\\
  Institute for Software Technology\\
  University Koblenz-Landau, Campus Koblenz}

\begin{document}

\maketitle

\begin{abstract}
  In Software Reengineering, one of the central artifacts is the source code of
  the legacy system in question.  In fact, in most cases it is the only
  definitive artifact, because over the time the code has diverged from the
  original architecture and design documents.  The first task of any
  reengineering project is to gather an understanding of the system's
  architecture.  Therefore, a common approach is to use parsers to translate
  the source code into a model conforming to the abstract syntax of the
  programming language the system is implemented in which can then be subject
  to querying.  Despite querying, transformations can be used to generate more
  abstract views on the system's architecture.

  This transformation case deals with the creation of a state machine model out
  of a Java syntax graph.  It is derived from a task that originates from a
  real reengineering project.
\end{abstract}

\section{Objective and Context}
\label{sec:objective}

The objective of the reengineering case presented here is to transform an
abstract Java syntax graph into a simple state machine.  There are two major
challenges involved.  The first challenge is an issue of \emph{performance} and
\emph{scalability}, because the input models are naturally large.

The second and more important challenge is that the transformation task
involves \emph{complex, non-local matching} of elements.  For example, the core
task, demands that the transformation should create one state for each Java
class that derives directly or indirectly from an abstract Java class named
``State.''  There are no restrictions on the depth of the inheritance
hierarchy, so the ``State'' class and its subclasses may be located arbitrarily
far in the input model.  However, the structure of the path from subclass to
superclass is clearly specified by the input metamodel and must be utilized by
transformations.

The SOAMIG\footnote{\url{http://www.soamig.de}} project dealt with the
migration of legacy systems to Service-Oriented Architectures by means of
model-driven techniques.  One legacy system on which the approach has been
evaluated is a monolithic Java system, which is operated with a graphical user
interface.  This user interface consists of around 30 different masks, which
often relate to conceptually self-contained functionalities that might be
implemented as services in the reengineered target system.  The order in which
masks are activated and which successor masks can be activated from a given
mask gives good hints about the orchestration of the target system services.

The masks are implemented as plain Java classes using the Swing toolkit.  Many
masks are very complex with many user interface elements and even more input
validation code, which complicates tracking down the relationships between the
individual masks.  However, the user interface is based on a state machine
concept and uses strict coding conventions in the implementation.  As such, any
masks can be seen as states, and when another mask is activated, it can be seen
as a transition.  The trigger of this transition is usually a click on some
button, and possibly additional actions are performed just before the
transition, e.g., validating user input.

A GReTL \cite{gretl-icmt2011} transformation has been developed which creates a
simple state machine model consisting of states and transitions with triggers
and actions out of the syntax graph of the legacy system consisting of more
than 2.5 million nodes and edges.  The transformation exploits the coding
conventions taken as a basis for the implementation of the graphical user
interface.  The resulting state machine model contains all information about
the possible sequences in which masks can be activated, what triggers are
responsible for a transition, and what additional actions are performed when
transitioning.  However, it consists of less than 100 nodes and edges, it can
be visualized and printed.  Therefore, it is of great value for the
understanding of the legacy system.

The transformation case proposed here is derived form this reengineering
project's task.  Instead of using the syntax graph of the proprietary system, a
toy example implementing the well-known TCP protocol state machine using very
similar coding conventions is used.  The next section describes the tasks
including the relevant metamodels and models.  The evaluation criteria used to
judge the solutions are discussed in Section~\ref{sec:evaluation-criteria}.

\section{Detailed Task Description}
\label{sec:task-descr}

In this section, the transformation task is explained in details.  The overall
goal of this task is to create a very simple \emph{state machine model} for a
\emph{Java syntax graph model} encoding a state machine with a set of coding
conventions a transformation has to exploit.

The task is divided into one mandatory \emph{core task} and two optional
\emph{extension tasks} with slightly increased complexity.  The conventions
used to implement the state machine in Java that should be exploited by
transformations are explained in terms of concrete Java syntax, i.e., by using
source code examples.  However, the most relevant metamodel elements are named,
too.  In the following, source metamodel elements are written underlined (e.g.,
\smme{MethodCall}), and target metamodel elements are written with a typewriter
font (e.g., \tmme{Transition}).

\paragraph{Source Metamodel and Models.}

The primary source model of the transformation is a Java abstract syntax graph
conforming to the JaMoPP Ecore metamodel \cite{jamopp09a,jamopp09b}.  The input
model contains any information present in the source code.  It consists of
about 6,500 elements.  A second input model is slightly larger and more
complex, e.g., the specialization hierarchy is deeper and so is the nesting of
statements in method bodies.  Lastly, a third industry-size input model is
provided, which was generated by parsing the source code of a Java project
containing about 900 classes and 220,000 lines of code resulting in a model
constisting of nearly one million elements.

All provided input models implement the TCP state machine according to the
conventions specified in the task description below, so the target state
machine model is always the same (not accounting for the order of elements).

\paragraph{Target Metamodel.}

\begin{figure}[h!]
  \centering
  \includegraphics[width=0.5\linewidth]{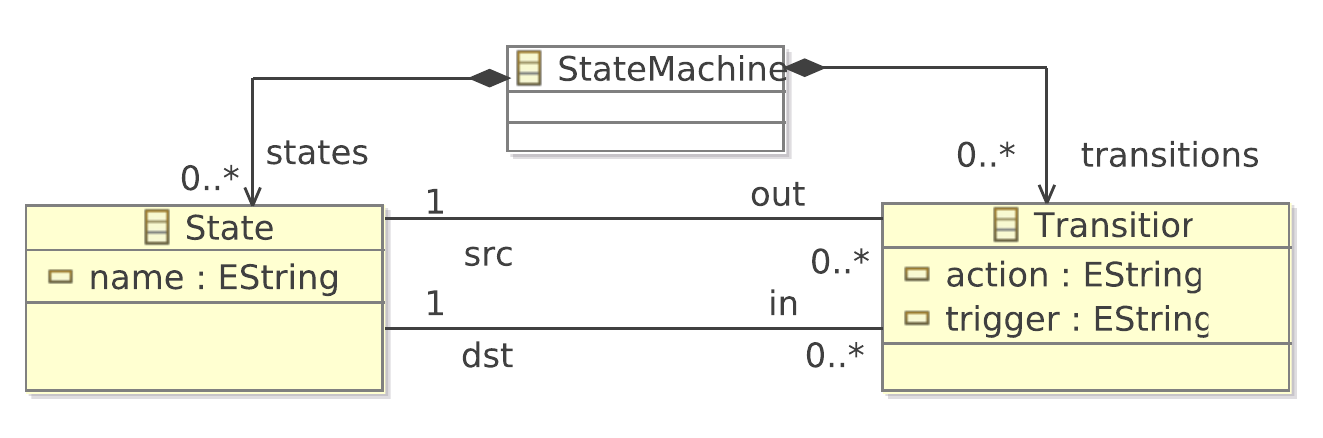}
  \caption{The target Ecore metamodel}
  \label{fig:state-machine-mm}
\end{figure}

As target metamodel, the very basic state machine metamodel shown in
Figure~\ref{fig:state-machine-mm} is used.
A \tmme{StateMachine} consists of an arbitrary number of \tmme{States} and
\tmme{Transitions}.  Any \tmme{Transition} starts at exactly one
\tmme{src}-\tmme{State}, and it ends at exactly one \tmme{dst}-\tmme{State}.
Every \tmme{State} has a \tmme{name}, and any \tmme{Transition} may be caused
by a \tmme{trigger}, and as a result of its activation an \tmme{action} might
be performed.

\paragraph{Core Task.}
\label{sec:core-task}

The core task should create a state machine model that contains all entities,
i.e., all \tmme{States} and all \tmme{Transitions} with the appropriate
references set.  Additionally, the \tmme{name} attribute defined for the
\tmme{State} class must be set.  The initialization of the \tmme{trigger} and
\tmme{action} attributes are left for the extension tasks.

Below, the coding conventions used to implement the TCP state machine in
plain-java are discussed in terms of concrete Java syntax and by using the
\lstinline{SynSent} class contained in the \texttt{src} directory, which is
shown in Listing~\ref{lst:syn-sent}.

\begin{lstlisting}[language=Java, caption={The \lstinline{SynSent} class}, label={lst:syn-sent}, float={h!t}]
public class SynSent extends ListeningState {
  private static State instance = new SynSent();
        public static State Instance() { return instance; }
        public void close() { Closed.Instance().activate(); }
        protected void run() {
                switch (getReceivedFlag()) {
                case SYN: send(Flag.SYN_ACK);
                          SynReceived.Instance().activate();
                          return;
                case SYN_ACK: send(Flag.ACK);
                              Established.Instance().activate();
                              return;
} } }
\end{lstlisting}

The coding conventions relevant for the core task are as follows:

\begin{compactenum}
\item A \lstinline{State} is a non-abstract Java class
  (\smme{classifiers.Class}) that extends the abstract class named ``State''
  directly or indirectly.  All concrete state classes are implemented as
  singletons \cite{DesignPatterns95}.  In Listing~\ref{lst:syn-sent},
  \lstinline{SynSent} extends the abstract \lstinline{ListeningState} state
  class, and that in turn extends the abstract \lstinline{State} class.
\item A \lstinline{Transition} is encoded by a method call
  (\smme{references.MethodCall}), which invokes the next state's
  \lstinline{Instance()} method (\smme{members.Method}) returning the singleton
  instance of that state on which the \lstinline{activate()} method is called
  in turn.  This activation may be contained in any of the classes' methods
  with an arbitrary deep nesting.  It may be assumed that a transition always
  has the form \lstinline{NewState.Instance().activate()}.  In the example,
  there are three transitions.  In the \lstinline{close()} method, there is one
  transition from the current state (\lstinline{SynSent}) into the
  \lstinline{Closed} state (line 7).  In the \lstinline{run()} method, there
  are another two transitions.  In line 11, there is a transition into the
  \lstinline{SynReceived} state, and in line 14, there is a transition into the
  \lstinline{Established} state.
\end{compactenum}

The target model state names should be set according to the Java classes they
were created for.  The outcome of the core task is a state machine with 11
states and 21 transitions between the states.

\paragraph{Extension 1: Triggers.}
\label{sec:ext1-triggers}

This extension task deals with the \lstinline{trigger} attribute of
transitions.  There are three different coding conventions that a
transformation has to exploit to set the correct trigger value.  These three
conventions and one fallback rule are specified as follows.

\begin{compactenum}
\item If the transition occurs in any method except \lstinline{run()}, then
  that method's name (\smme{members.Method.name}) shall be used as the trigger.
  For example, the activation of \lstinline{Closed} in line 7 of
  Listing~\ref{lst:syn-sent} occurs in the \lstinline{close()} method, so the
  trigger is \lstinline{close}.
\item If the transition occurs inside a non-default \lstinline{case} block
  (\smme{statements.NormalSwitchCase}) of a \lstinline{switch} statement
  (\smme{statements.Switch}) in the \lstinline{run()} method, then the
  enumeration constant (\smme{members.EnumConstant}) used as condition of the
  corresponding \lstinline{case} is the trigger.  For example, when activating
  \lstinline{SynReceived} in line 11 of Listing~\ref{lst:syn-sent}, the trigger
  is \lstinline{SYN}.
\item If the transition occurs inside a \lstinline{catch} block
  (\smme{statements.CatchBlock}) inside the \lstinline{run()} method, then the
  trigger is the name of the caught exception's class.
\item If none of the three cases above apply, i.e., the activation call is
  inside the \lstinline{run()} method but without a surrounding
  \lstinline{switch} or \lstinline{catch}, the corresponding transition is
  triggered unconditionally.  In that case, the trigger attribute shall be set
  to \lstinline{--}.
\end{compactenum}
Transformations may assume that the four different cases can be matched without
ambiguity, e.g., there is no \lstinline{catch} block activating some state
inside a surrounding \lstinline{switch}, or vice versa.

\paragraph{Extension 2: Actions.}
\label{sec:ext2-actions}

The task of the second extension is to set the \lstinline{action} attributes of
transitions.  The action of a transition is specified as follows.

\begin{compactenum}
\item If the block (\smme{statements.StatementListContainer}) containing the
  transition to the next state contains a method call to the \lstinline{send()}
  method, then that call's enumeration constant parameter's name is the action.
  For example, the \lstinline{action} attribute of the transition in line 11 of
  Listing~\ref{lst:syn-sent} has to be set to \lstinline{SYN_ACK}.
\item If there is no call to \lstinline{send()} in the activation call's block,
  the \lstinline{action} of the corresponding transition shall be set to
  \lstinline{--}.  For example, the transition in line 7 of
  Listing~\ref{lst:syn-sent} performs no action, and thus the \textsf{action}
  has to be set to \lstinline{--}.
\end{compactenum}

\begin{figure}[h!]
  \centering
  \includegraphics[width=0.8\linewidth]{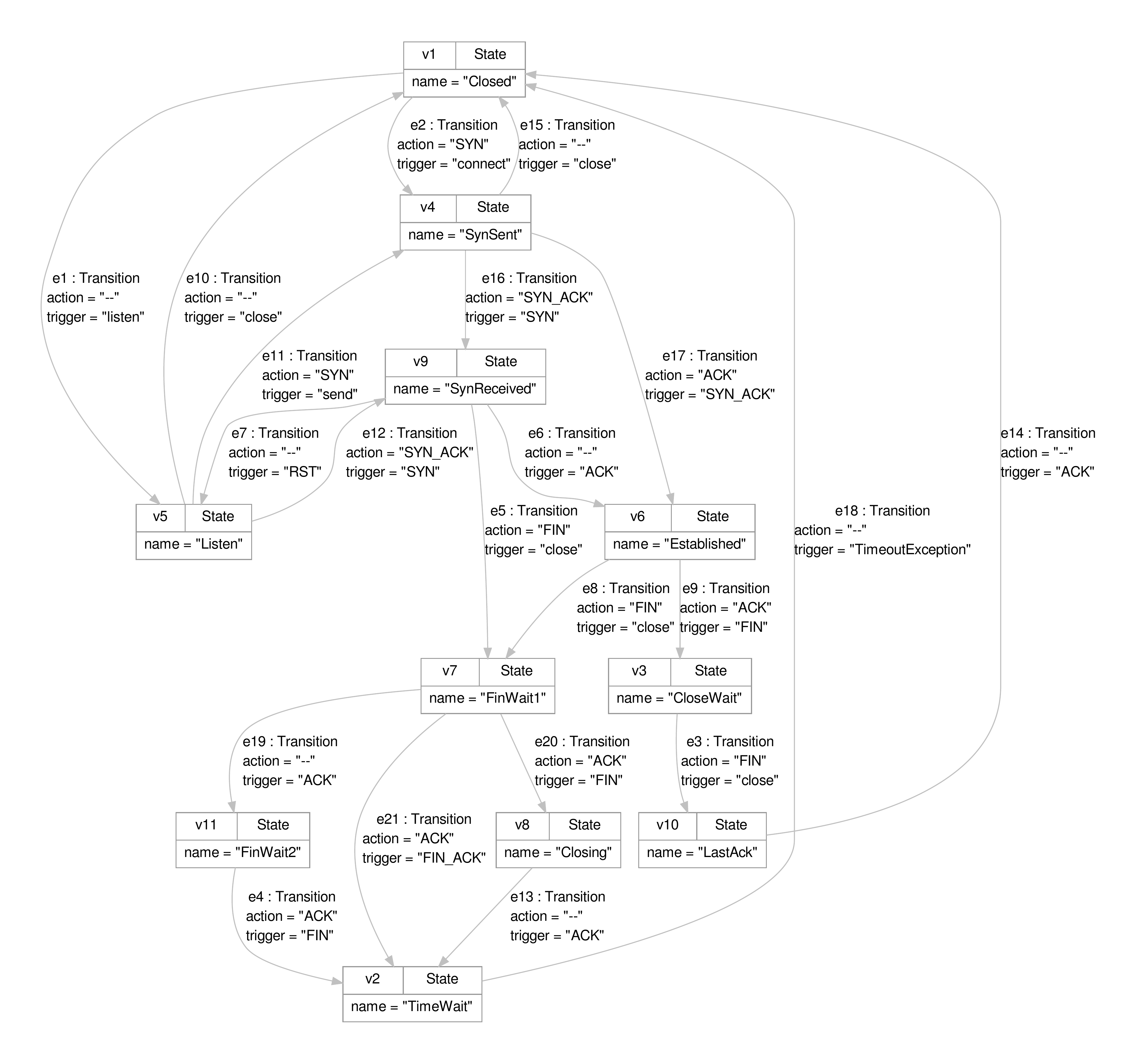}
  \caption{The final state machine after performing the core and both extension
    tasks}
  \label{fig:ext2-task-result}
\end{figure}

A visualization of the target state machine produced by the reference solution
is shown in Figure~\ref{fig:ext2-task-result}.

\section{Evaluation Criteria}
\label{sec:evaluation-criteria}

As motivated in Section~\ref{sec:objective}, the goal of this transformation
case is supporting program understanding.  By facilitating a set of coding
conventions, model transformations can be used to accomplish the task of
extracting the implicitly encoded state machine, in contrast to modeling it
manually by thoroughly inspecting all relevant classes of the system in
question.  In order to have a feasible solution, the time needed for writing
and executing the transformation must be comparable to the time that would be
needed for a code inspection and manually modelling the state machine.
However, if we assume that the set of coding conventions derived from the
initial brief inspection is correct, it can be assumed that the transformation
produces a correct output without human mistakes.

Since the speed of writing a solution cannot be judged directly,
\textbf{understandability} and \textbf{conciseness} are used as objectively
ascertainable measures relating to the implementation effort, weighted with
30\%.  Ideally, each coding convention described in
Section~\ref{sec:task-descr} results in a transformation rule in which the
statement of the convention is clearly visible.

The \textbf{correctness} of the solution is also important.  If the model
created by the transformation is the foundation of weighty decisions, it would
be fatal if it didn't reflect the reality.  The \textbf{completeness} of a
solution is closely entangled to correctness, because an incomplete model may
also lead to false decisions.  Both correctness and completeness are weighted
with 17.5\%.

The last property that will be judged is the \textbf{performance}, weighted
with 5\%.  It is much less important than the other criteria, but of course
such a transformation should be applicable on common hardware in acceptable
time.

\bibliography{case}

\begin{thebibliography}{1}
\providecommand{\bibitemdeclare}[2]{}
\providecommand{\urlprefix}{Available at }
\providecommand{\url}[1]{\texttt{#1}}
\providecommand{\href}[2]{\texttt{#2}}
\providecommand{\urlalt}[2]{\href{#1}{#2}}
\providecommand{\doi}[1]{doi:\urlalt{http://dx.doi.org/#1}{#1}}
\providecommand{\bibinfo}[2]{#2}

\bibitemdeclare{book}{DesignPatterns95}
\bibitem{DesignPatterns95}
\bibinfo{author}{Erich Gamma}, \bibinfo{author}{Richard Helm},
  \bibinfo{author}{Ralph Johnson} \& \bibinfo{author}{John Vlissides}
  (\bibinfo{year}{1995}): \emph{\bibinfo{title}{{Design Patterns}}}.
\newblock \bibinfo{publisher}{Addison-Wesley}.

\bibitemdeclare{incollection}{jamopp09a}
\bibitem{jamopp09a}
\bibinfo{author}{Florian Heidenreich}, \bibinfo{author}{Jendrik Johannes},
  \bibinfo{author}{Mirko Seifert} \& \bibinfo{author}{Christian Wende}
  (\bibinfo{year}{2010}): \emph{\bibinfo{title}{{Closing the Gap between
  Modelling and Java}}}.
\newblock In \bibinfo{editor}{M.~van~den Brand}, \bibinfo{editor}{D.~Gasevic}
  \& \bibinfo{editor}{J.~Gray}, editors: {\sl \bibinfo{booktitle}{Software
  Language Engineering}}, {\sl \bibinfo{series}{Lecture Notes in Computer
  Science}} \bibinfo{volume}{5969}, \bibinfo{publisher}{Springer}, pp.
  \bibinfo{pages}{374--383}, \doi{10.1007/978-3-642-12107-4\_25}.

\bibitemdeclare{techreport}{jamopp09b}
\bibitem{jamopp09b}
\bibinfo{author}{Florian Heidenreich} et~al. (\bibinfo{year}{2009}):
  \emph{\bibinfo{title}{{JaMoPP: The Java Model Parser and Printer}}}.
\newblock \bibinfo{type}{Technical Report} \bibinfo{number}{TUD-FI09-10},
  \bibinfo{institution}{Technische Universit\"at Dresden, Fakult\"at
  Informatik}.

\bibitemdeclare{inproceedings}{gretl-icmt2011}
\bibitem{gretl-icmt2011}
\bibinfo{author}{Tassilo Horn} \& \bibinfo{author}{J\"urgen Ebert}
  (\bibinfo{year}{2011}): \emph{\bibinfo{title}{The GReTL Transformation
  Language}}.
\newblock In \bibinfo{editor}{Jordi Cabot} \& \bibinfo{editor}{Eelco Visser},
  editors: {\sl \bibinfo{booktitle}{Theory and Practice of Model
  Transformations, Fourth International Conference, ICMT 2011, Zurich,
  Switzerland, June 27-28, 2011. Proceedings}}, {\sl \bibinfo{series}{Lecture
  Notes in Computer Science}} \bibinfo{volume}{6707},
  \bibinfo{publisher}{Springer}, pp. \bibinfo{pages}{183--197},
  \doi{10.1007/978-3-642-21732-6\_13}.

\end{thebibliography}
\bibliographystyle{eptcs}

\end{document}